\documentclass[prx,aps,twocolumn,showpacs,eqsecnum,superscriptaddress]{revtex4-1}
\usepackage{amssymb}
\usepackage{amsmath}
\usepackage{graphicx}
\usepackage{wasysym}
\usepackage{color}
\usepackage{mathtools}

\newcommand{\be}{\begin{eqnarray}}
\newcommand{\ee}{\end{eqnarray}}
\newcommand{\bbm}{\begin{bmatrix}}
\newcommand{\ebm}{\end{bmatrix}}
\newcommand{\bpm}{\begin{pmatrix}}
\newcommand{\epm}{\end{pmatrix}}
\renewcommand{\v}[1]{{\bf #1}}
\newcommand{\nn}{\nonumber \\}

\newcommand{\bp}{\bar{p}}
\newcommand{\bpz}{\bar{p}_z}
\newcommand{\bpr}{\bar{p}_r}
\newcommand{\br}{\bar{r}}
\newcommand{\brho}{\bar{\rho}}
\newcommand{\bz}{\bar{z}}

\begin{document}
\title[]{Anisotropic density fluctuations, plasmons, and Friedel oscillations in nodal line semimetal}

\author{Jun-Won \surname{Rhim}}
\affiliation{Max-Planck-Institut f{\"u}r Physik komplexer Systeme, 01187 Dresden, Germany}

\author{Yong Baek \surname{Kim}}
\affiliation{Department of Physics and Center for Quantum Materials,
University of Toronto, Toronto, Ontario M5S 1A7, Canada}
\affiliation{Canadian Institute for Advanced Research, Toronto, Ontario, M5G 1Z8, Canada}


\begin{abstract}
Motivated by recent experimental efforts on three-dimensional semimetals, we investigate the static and dynamic density response of the nodal line semimetal by computing the polarizability for both undoped and doped cases. The nodal line semimetal in the absence of doping is characterized by a ring-shape zero energy contour in momentum space, which may be considered as a collection of Dirac points. In the doped case, the Fermi surface has a torus shape and two independent processes of the momentum transfer contribute to the singular features of the polarizability even though we only have a single Fermi surface. In the static limit, there exist two independent singularities in the second derivative of the static polarizability. This results in the highly anisotropic Friedel oscillations which show the angle-dependent algebraic power law and the beat phenomena in the oscillatory electron density near a charged impurity. Furthermore, the dynamical polarizability has two singular lines along $\hbar\omega = \gamma p$ and $\hbar\omega = \gamma p \sin\eta$, where $\eta$ is the angle between the external momentum $\v p$ and the plane where the nodal ring lies. 
From the dynamical polarizability, we obtain the plasmon modes in the doped case, which show anisotropic dispersions and angle-dependent plasma frequencies. Qualitative differences between the low and high doping regimes are discussed in light of future experiments. 
\end{abstract}



\maketitle


\section{Introduction}

Recently, three-dimensional (3D) semimetals with unusual band topology have received great attention due to their novel physical properties.
The most widely studied example is the Weyl semimetal which hosts pairs of massless Weyl fermions at the nodal points in the 3D bulk Brillouin zone \cite{Murakami,Wan,Balents,Burukov, Vazifeh, Huang1, Xu1, Lv1, Xu2, Yang1, Lv2}.
A pair of Weyl points with opposite chirality act as monopole and anti-monopole of the Berry curvature, which leads to the surface Fermi arc 
that connects the projections of a pair of Weyl points in the surface Brillouin zone \cite{Murakami,Wan,Balents,Burukov,Vazifeh}.
Various novel physical properties are predicted to occur or have been observed, including the negative magnetoresistance associated with the chiral anomaly\cite{Zyuzin, Jan, Huanf2, Yang2}, the pressure induced anomalous Hall effect\cite{Yang3}, and the intriguing quantum oscillations\cite{Potter}. 
Another interesting example is the quadratic band touching semimetal which also has a nodal point as a Fermi surface, but with the parabolic conduction and valence bands around the touching point \cite{Bohm, Krempa}.
Its peculiar band structure is reflected in the aperiodic quantum oscillation\cite{Rhim1}.
When the long-range Coulomb interaction is taken into account\cite{Moon,Herbut}, a quantum critical non-Fermi-liquid phase is expected to arise\cite{Moon}.

In contrast to the semimetals with the point nodes, the nodal line semimetal in three-dimensions, which is the subject of this paper, is characterized by a one-dimensional ring-shaped line node at zero energy. 
The nodal line spectra can be found in a variety of systems and they are protected by various discrete symmetries such as the time-reversal, mirror, inversion or nonsymmorphic symmetries, depending on the materials\cite{Burukov, Carter, Phillips, Chen, HSKim, YKim, Xie, Zeng, Weng1, Yu, Weng2, Sekine, Chan, Fang, Schaffer,Lee,Mullen}.
When the particle-hole symmetry exists, a flat band arises in some portions of the surface Brillouin zone, which is a result of topological properties of
the nodal ring spectrum in the bulk. The nodal line can also be regarded as an infinite collection of the 2D Dirac dispersions, each defined 
in the 2D planes in momentum space, which intersect the ring in perpendicular directions.
It was shown that 3D almost-flat Landau levels appear when the magnetic field is applied along the parallel direction to the ring and 
this may manifest as a 3D quantum Hall effect\cite{Rhim2, Koshino}.
Furthermore, when the Coulomb interaction is included, the system has a non-trivial interacting fixed point\cite{Huh}.
Possible superconductivity in such systems is also investigated\cite{nandkishore}.

In this paper, we investigate the dynamical density response of the nodal line semimetal for doped and undoped cases by computing the dynamical polarizability
in random phase approximation (RPA). In general, the static and dynamical polarizability are highly anisotropic and depend strongly on the direction of the momentum transfer.
When the nodal line semimetal is doped, the Fermi surface has a torus shape in momentum space. This allows two characteristic momentum-transfer processes that lead to two direction-dependent singularities in the polarizability. It is shown that these singularities are responsible for the anisotropic Friedel oscillations of the induced 
electron density near a charged impurity. Furthermore, the plasmon dispersion and plasma frequency become highly anisotropic and angle-dependent.

The rest of the paper is organized as follows.
In Section II, we introduce the minimal low energy Hamiltonian for the nodal line semimetal. We also elucidate the relation between the
small momentum transfer processes in 3D nodal line semimetal and those of 2D Dirac fermion systems.
In Section III, we compute the bare polarizability.
For the neutral case, we obtain fully analytic expressions.
Although the complete analytic description is not accessible for the doped case, we derive relevant approximate analytic formulae in various important limits.
In Section IV, the anisotropic plasmon modes with unusual density dependence for the doped nodal line semimetal are studied based on the RPA polarizability.
In Section V, the Friedel oscillations of the nodal line semimetal induced by a charged impurity are analyzed.
The induced charge density oscillations exhibit a number of characteristic behaviors such as the anisotropic algebraic power law and beat phenomena 
in the oscillations, which are related to the double singularity structure in the static polarizability.
We conclude in Section VI.

\section{The continuum model for the nodal line semimetal}
\begin{figure}
\includegraphics[width=0.8\columnwidth]{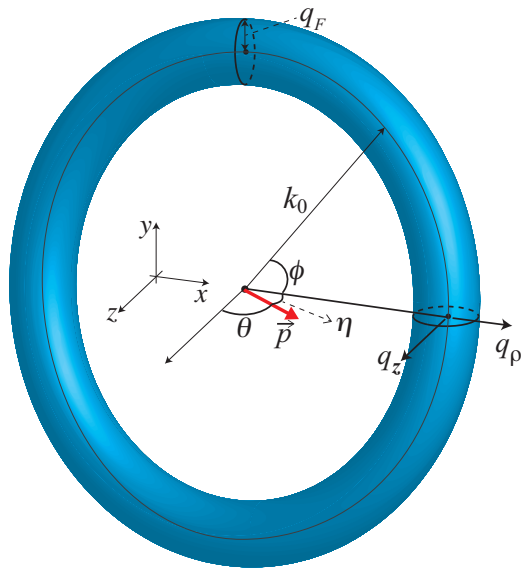}
\caption{(Color online) {The toroidal coordinate for the nodal line semimetal. The nodal line is at $k_r=k_0$. The minor radius $q_F$ is the Fermi momentum. $\v p$ is the external momentum which has an angle $\eta$ with the $xy$ plane and is assumed to have no $y$ component without loss of generality. Around each nodal point, a local 2D cartesian coordinate is described by $q_r$ and $q_z$.} }
\label{fig:coordinate}
\end{figure}

The minimal low energy Hamiltonian for the nodal line semimetal is given by
\be
H_{R} = \gamma q_\rho \tau_x + \gamma q_z \tau_y \label{eq:ring_ham},
\ee
where $\tau_x$ and $\tau_y$ are the Pauli matrices.
The Hamiltonian is described in the toroidal coordinate system, where $q_\rho$ and $q_z$ are the momentum components along the radial and $z$ direction as shown in Fig. \ref{fig:coordinate}.
In the original cartesian coordinate, they read $q_\rho  =\sqrt{k_x^2+k_y^2} -k_0 = k_\rho-k_0$ and $q_z = k_z$, where $k_0$ is the radius of the nodal ring.
The energy spectrum of (\ref{eq:ring_ham}) is evaluated as 
\be
E_s(\v k) = \varepsilon_s(\phi,\v q) = s\gamma\sqrt{q_\rho^2+q_z^2} \label{eq:ring_energy}
\ee
with the band index $s=\pm 1$. 
It has zero modes on the Fermi circle described by $k_z^2+k_y^2=k_0^2$ and $k_z=0$.
The Hamiltonian of the nodal line semimetal can be regarded as an infinite collection of the Dirac Hamiltonians defined at every polar angle $\phi$.
This is emphasized by introducing another notation $\varepsilon_s(\phi,\v q)$ in the above which represents the energy dispersion of graphene at each polar angle $\phi$.

For later use, we consider a momentum transfer process, which leads to the change in dispersion $E_s(\v k) \rightarrow E_s(\v k + \v p)$, 
where $\v p$ will be called the \textit{external momentum}.
Since the nodal ring is isotropic, we assume without loss of generality that $\v p = p\cos\eta \hat{x} + p\sin\eta\hat{z}$ where $\eta$ is the angle of the external momentum ${\v p}$ from the plane in which the nodal ring lies ($xy$-plane in Fig. \ref{fig:coordinate}).
If the size of the nodal ring is much larger than the momentum transfer ($k\approx k_0 \gg |p|$), we have
\be
E_s(\v k + \v p) \approx \varepsilon_s(\phi,\v q + \v p^\prime) \label{eq:ring_energy_transfer}
\ee
where 
\be
\v p^\prime = p\cos\phi\cos\eta\hat{\rho} + p\sin\eta\hat{z} \label{eq:p_prime}
\ee
since $\v k + \v p \big|_\rho \approx k_\rho + p\cos\phi\cos\eta$ and $\v k + \v p \big|_z = k_z + p\sin\eta$.
In other words, one can calculate the energy at the shifted momentum $\v k + \v p$ from the energy of graphene at the polar angle $\phi$ with the momentum transfer $\v p^\prime$ in the toroidal coordinate.
The most advantageous thing in this approximation is that $\v p^\prime$, the external momentum in the toroidal coordinate, is independent of $\v q$ so that one can utilize various results obtained for graphene.

The relevance of the Coulomb interaction in this system can be examined as follows.
For massless particles, the Wigner-Seitz radius is defined as $r_s = \langle V \rangle / E_F$ which measures the ratio between the Coulomb and kinetic energies.
In the case of graphene, both $\langle V \rangle$ and $E_F$ are proportional to $q_F$, the Fermi momentum.
As a result, $r_s$ can be smaller than 1 for both undoped and doped cases if it is on the appropriate substrates.\cite{Sarma}
For the nodal line semimetal, on the other hand, one can show that $r_s = a_0 (k_0 / q_F)^{1/3}$ where $a_0 = (g/3)^{1/3}e^2/\kappa_0\gamma$ (dimensionless) because $n_e = 1/(\frac{4}{3}\pi\langle r\rangle^3) = (g/4\pi)k_0q_F^2$ and $\langle V \rangle = e^2/\kappa_0\langle r\rangle$.
Here, $g$ is the degeneracy or the number of degenerate nodal lines, $\kappa_0$ is the vacuum permeability, 
and $n_e$ is the electron density.
Then the condition $r_s \ll 1$ becomes
\be
\frac{k_0}{q_F} \ll \left( \frac{\kappa_0\gamma}{e^2} \right)^3 \frac{3}{g}. \label{eq:valid}
\ee 
This means that any perturbative analysis would not be valid for the neutral case ($q_F =0$).
For doped cases, on the other hand, one can fulfill the above condition relatively easily  due to the cubic power on the righthand side.


\begin{figure*}
\includegraphics[width=2\columnwidth]{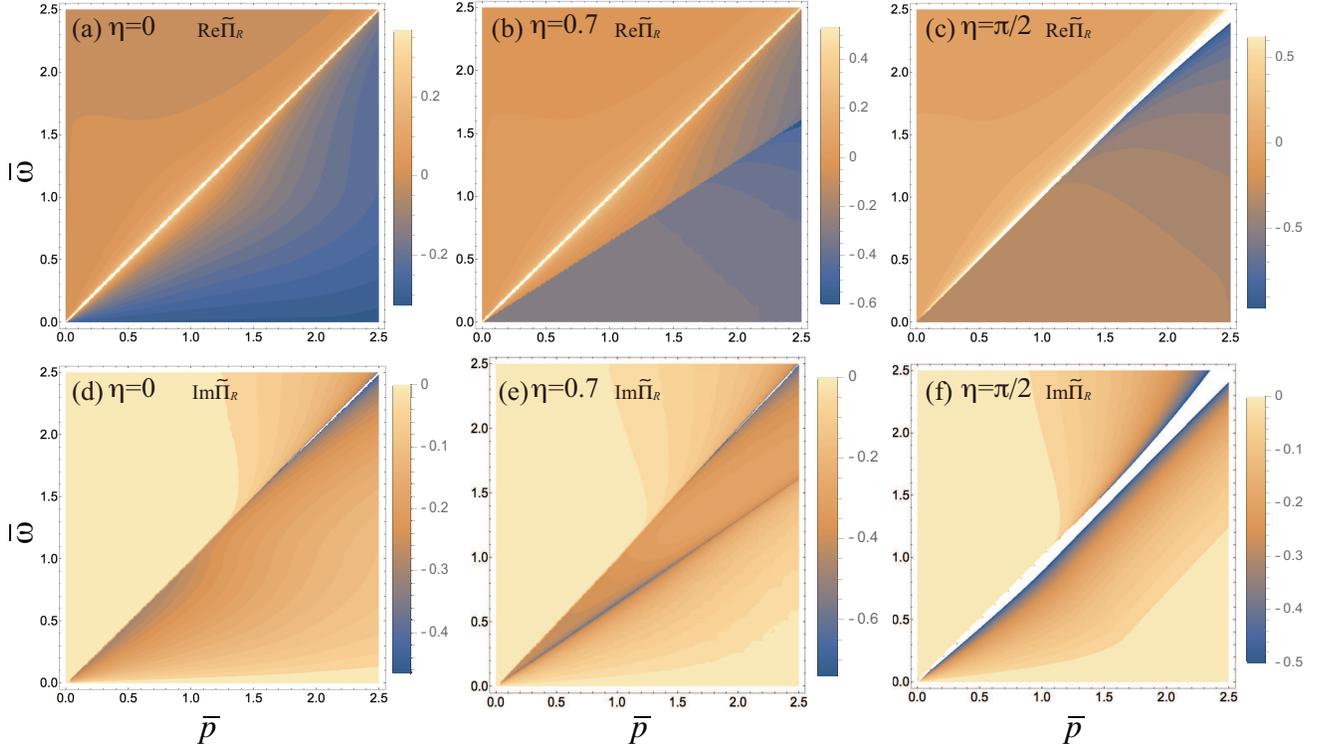}
\caption{(Color online) Contour plots of the real and imaginary parts of the polarizability of the nodal line semimetal for various values of $\eta$. The regions where the magnitudes starts to increase rapidly remain uncolored.}
\label{fig:pol_0}
\end{figure*}

\section{Polarizability}
The polarizability of the nodal line semimetal is defined by
\be
\Pi_R(\v p,\omega) = \frac{g}{8\pi^3}\sum_{s,s^\prime,\v k} S^{s,s^\prime}_{\v k, \v k^\prime} \frac{f_{s,\v k} - f_{s^\prime,\v k^\prime}}{\hbar\omega+E_{s,\v k}-E_{s^\prime,\v k^\prime} +i\epsilon} \label{eq:pol_def}
\ee
where $f_{s,\v k}$ is the Fermi-Dirac distribution and $S^{s,s^\prime}_{\v k, \v k^\prime}$ is a structure factor determined by the overlap between wavefunctions at $\v k$ in the $s$-th band and $\v k^\prime = \v k +\v p$ in the $s^\prime$-th band.\cite{Fetter}
Within the large ring approximation, one can just apply the relationships (\ref{eq:ring_energy}) and (\ref{eq:ring_energy_transfer}) to (\ref{eq:pol_def}).
Also, one can use the structure factor of graphene, $S^{s,s^\prime}_{\v k, \v k^\prime} = (1+ss^\prime\cos\phi_{\v q,\v q +\v p^\prime})/2$, for this case since the Hamiltonian (\ref{eq:ring_ham}) at each polar angle is exactly the same as that of graphene.
Then, the polarizability of the nodal line semimetal is given by
\be
\Pi_R(\v p,\omega) = \frac{gk_0}{2\pi}\int_0^{2\pi} d\phi \Pi_{G}(\v p^\prime,\omega) \label{eq:ring_pol_0}
\ee
where $\Pi_{G}(\v p^\prime,\omega)$ is the polarizability of graphene.
Note that $\v p^\prime$ depends on the polar angle $\phi$ as shown in (\ref{eq:p_prime}).
The polarizability of graphene is usually decomposed into two parts as $\Pi_{G}(\v p,\omega) = \Pi^-_{G}(\v p,\omega) + \Pi^+_{G}(\v p,\omega)$,
which are given by
\be
\Pi^-_{G}(\v p,\omega) = -i\frac{g}{16}\frac{p^2}{\sqrt{\hbar^2\omega^2 - \gamma^2 p^2}}
\ee
and
\be
\Pi^+_{G}(\v p,\omega) &=& -\frac{\mu}{\gamma^2}\frac{g}{2\pi} + \frac{g}{16\pi}\frac{p^2}{\sqrt{\hbar^2\omega^2 - \gamma^2 p^2}} \Big[ G(u_{p,\omega}) \nn
&& -\Theta(u_{p,-\omega}-1)\left\{G(u_{p,-\omega}) -i\pi\right\} \nn
&& -\Theta(1-u_{p,-\omega})G(-u_{p,-\omega}) \Big] \label{eq:grp_pol_imag}
\ee
where $u_{p,\omega} = (\hbar\omega + 2\mu)/\gamma p$ and $G(x) = x\sqrt{x^2-1}-\ln(x+\sqrt{x^2-1})$.\cite{Hwang,Min,Wunsch,Ando}
For the undoped case ($\mu =0$), $\Pi^+_{G}(\v p,\omega) $ vanishes.

\subsection{Undoped case}
As in the case of graphene, we define the neutral part of the polarizability of the nodal line semimetal as $\Pi^-_R(\v p,\omega) = (gk_0/2\pi)\int_0^{2\pi} d\phi \Pi^-_{G}(\v p^\prime,\omega)$.
With the replacement of $p^\prime = p\sqrt{\cos^2\eta \cos^2\phi + \sin^2\eta}$, we obtain the real and imaginary parts of the polarizability as follows:
\be
\mathrm{Re}\Pi^-_R(\v p,\omega) &=& -\lim_{\mu\rightarrow 0} \frac{\mu k_0}{\gamma^2}\left[ \Lambda^{(r)}_E(\bar{p},\bar{\omega}) + \Lambda^{(r)}_F(\bar{p},\bar{\omega})\right] \label{eq:real_pol_neutral}\nn
&& \times \Theta(\bar{p} - \bar{\omega})
\ee
and
\be
\mathrm{Im}\Pi^-_R(\v p,\omega) &=& \lim_{\mu\rightarrow 0} \frac{\mu k_0}{\gamma^2}\left[ \Lambda^{(i)}_E(\bar{p},\bar{\omega}) - \Lambda^{(i)}_F(\bar{p},\bar{\omega})\right]\label{eq:imag_pol_neutral}\nn
&& \times \Theta(\bar{\omega}-\bar{p}\sin\eta)
\ee
where
\be
\Lambda^{(r)}_E(\bar{p},\bar{\omega}) &=& \frac{g}{8\pi}\sqrt{\bar{p}^2-\bar{\omega}^2} E\left(\phi_r,\frac{\bar{p}^2\cos^2\eta}{\bar{p}^2-\bar{\omega}^2}\right),\\
\Lambda^{(r)}_F(\bar{p},\bar{\omega}) &=& \frac{g}{8\pi}\frac{\bar{\omega}^2}{\sqrt{\bar{p}^2-\bar{\omega}^2}}F\left(\phi_r,\frac{\bar{p}^2\cos^2\eta}{\bar{p}^2-\bar{\omega}^2}\right), \\
\Lambda^{(i)}_E(\bar{p},\bar{\omega}) &=& \frac{g}{8\pi}\sqrt{\bar{\omega}^2 -\bar{p}^2\sin^2\eta} E\left(\phi_i,\frac{\bar{p}^2\cos^2\eta}{\bar{\omega}^2 -\bar{p}^2\sin^2\eta}\right),\nn
\ee
and
\be
\Lambda^{(i)}_F(\bar{p},\bar{\omega}) &=& \frac{g}{8\pi}\frac{\bar{\omega
}^2}{\sqrt{\bar{\omega}^2 -\bar{p}^2\sin^2\eta}} F\left(\phi_i,\frac{\bar{p}^2\cos^2\eta}{\bar{\omega}^2 -\bar{p}^2\sin^2\eta}\right).\nn
\ee
Here, we define dimensionless parameters $\bar{p} = p/q_F = \gamma p/\mu$ and $\bar{\omega} = \hbar\omega / \mu$, where $\mu = \gamma q_F$ is the chemical potential.
Although $q_F = \mu = 0$ for the neutral case, we introduced those notations to have a unified description with the doped case. 
We just take the limit $\mu \rightarrow 0$ for the undoped case. 
Here, $F(\phi,x)$ and $E(\phi,x)$ are the elliptic integrals of the first and second kind.
The amplitudes of the elliptic integrals are given by
\be
\phi_r &=& \cos^{-1}\left(\frac{\sqrt{\bar{\omega}^2-\bar{p}^2\sin^2\eta}}{\bar{p}\cos\eta}\Theta(\bar{\omega}-\bar{p}\sin\eta)\right)~
\ee 
and
\be
\phi_i &=& \frac{\pi}{2} - \cos^{-1}\left(\frac{\sqrt{\bar{\omega}^2-\bar{p}^2\sin^2\eta}}{\bar{p}\cos\eta}\right)\Theta(\bar{p}-\bar{\omega})~.
\ee
Although the amplitudes in the above are not well-defined when $\bar{\omega}>\bar{p}$ for $\phi_r$ and $\bar{\omega}<\bar{p}\sin\eta$ for $\phi_i$, those intervals are already forbidden by the step functions in (\ref{eq:real_pol_neutral}) and (\ref{eq:imag_pol_neutral}).

Both the real and imaginary parts of the polarizability of the nodal line semimetal are non-vanishing when $\gamma p\sin\eta <\hbar\omega<\gamma p$.
Otherwise, it is real or purely imaginary as in the case of graphene.
If $\eta = \pi/2$ where the momentum transfer is in the direction perpendicular to the plane of the nodal ring, the behavior of the polarizability of the nodal ring semimetal is exactly the same as that of graphene since $p^\prime$ has no $\phi$-dependence in this case.
One can check that, when $\eta=\pi/2$, $\Pi_R(\v p,\omega) = gk_0 \Pi_{G}(\v p,\omega)$ because $E(0,0)=F(0,0)=\pi/2$ and $E(\pi/2,0)=F(\pi/2,0)=\pi/2$.

We examine the low-frequency properties of the the polarizability.
For finite $\eta$, $\mathrm{Im}\Pi^-_R(\v p,\omega)$ is just vanishing as seen from the step function in (\ref{eq:imag_pol_neutral}) and has no frequency dependence.
On the other hand, when the momentum transfer is along the radial direction ($\eta = 0$), the imaginary part of the polarizability becomes finite around $\bar{\omega} = 0$ and is found as
\be
\mathrm{Im}\Pi^-_R(\v p,\omega) \approx -\frac{\mu k_0}{\gamma^2}\frac{3g}{32\pi}\frac{\bar{\omega}^2}{\bar{p}}
\ee
for arbitrary momentum $\bar{p}$.
On the other hand, the real part of the polarizability has a logarithmic behavior in the low-frequency regime as
\be
\mathrm{Re}\Pi^-_R(\v p,\omega) \approx -\frac{\mu k_0}{\gamma^2}\frac{g}{8\pi}\left( \bar{p}-\frac{1}{2}\frac{\bar{\omega}^2}{\bar{p}}\ln\frac{e}{4}\frac{\bar{\omega}}{\bar{p}} \right)
\ee
when $\eta =0$.

\begin{table}
\begin{tabular}{l*{3}{c}}
\hline
             & $\quad \bar{p}<2 \quad $ & $\quad 2 \leq \bar{p} <\frac{2}{\sin\eta} \quad $ & $\quad \frac{2}{\sin\eta} < \bar{p} \quad $  \\
\hline
$\eta=0$     & $\bar{\omega} \ln\bar{\omega}$ & $\bar{\omega} \ln\bar{\omega}$ & -   \\
$0<\eta<\frac{\pi}{2}$            & $\bar{\omega}$ & $\bar{\omega}$ & 0   \\
$\eta=\frac{\pi}{2}$           & $\bar{\omega}$ & $\bar{\omega}^{\frac{3}{2}}$ & 0     \\
\hline
\end{tabular}
\caption{ Low-frequency behaviors of the imaginary part of the polarizability in the doped case, for various values of $\eta$. }
\label{table_1}
\end{table}

\subsection{Doped case}
For the doped cases, we should include both $\Pi^-_{G}(\v p,\omega) $ and $\Pi^+_{G}(\v p,\omega) $ in (\ref{eq:ring_pol_0}) for the evaluation of the polarizability of the nodal line semimetal.
Both real and imaginary parts of the polarizability are plotted in Fig. \ref{fig:pol_0} for various values of $\eta$.
As in the case of graphene, the polarizability of the nodal line semimetal also has a singular path along $\bar{\omega} = \bar{p}$ for both real and imaginary parts.
When $\bar{\omega} \approx \bar{p}$ and $\bar{p}<1$, the polarizability can be written as follows:
\be
\mathrm{Re}\tilde{\Pi}_R(\v p,\omega) \approx -\frac{g}{2\pi^2} \frac{\bar{\omega}}{\bar{p}\cos\eta}\ln\frac{1}{16}\frac{|\bar{p}^2-\bar{\omega}^2|}{\bp^2\cos^2\eta}-\frac{g}{2\pi}  
\ee
and
\be
\mathrm{Im}\tilde{\Pi}_R(\v p,\omega) &\approx & -\left[ \frac{g}{2\pi}\frac{\bar{\omega}}{\bar{p}\cos\eta} + \frac{g}{8\pi}\frac{\bar{\omega}(\bar{p}^2 - \bar{\omega}^2)}{\bar{p}^3 \cos^3\eta} \right]\nn
&& \times \Theta(\bar{p}-\bar{\omega})
\ee
where we define the dimensionless quantity, 
\be
\tilde{\Pi}_R(\v p,\omega) = \frac{\gamma^2}{\mu k_0}\Pi_R(\v p,\omega).
\ee
The real part of the polarizability has a logarithmic divergence along $\bar{\omega} = \bar{p}$ while the imaginary part is finite-valued albeit being discontinuous across the line $\bar{\omega} = \bar{p}$  when $\bar{p}<1$.
 However, the imaginary part of the polarizability also starts to diverge at $\bar{\omega} = \bar{p}$ when $\bar{p}$ is larger than unity.
Note that this approximation is valid only when $\eta < \pi/2$.
If $\eta = \pi/2$, the situation becomes exactly the same as the case of graphene and  both the real and imaginary parts of the polarizability for ${\bar p} < 1$ diverge along $\bar{\omega} = \bar{p}$.

\begin{figure*}
\includegraphics[width=2\columnwidth]{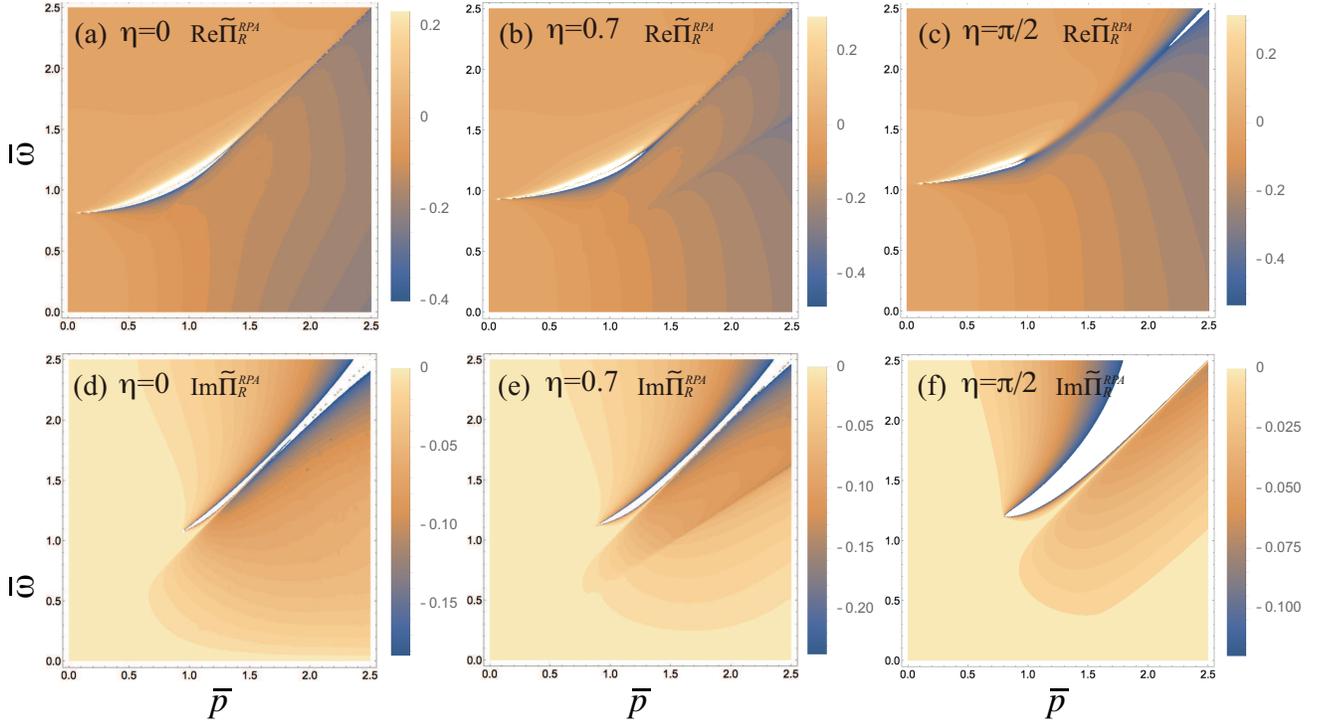}
\caption{(Color online) Contour plots of the real and imaginary parts of the renormalized polarizability of the nodal line semimetal for various values of $\eta$. The regions where the magnitude starts to increase rapidly remain uncolored. }
\label{fig:pol_rpa}
\end{figure*}

Another interesting feature of the polarizability of the nodal line semimetal is that we have an additional singular path along $\bar{\omega} = \bar{p}\sin\eta$ as shown in Fig. \ref{fig:pol_0}.
We obtain an approximate expression for the polarizability when $\bar{\omega} \approx \bar{p}\sin\eta$ and $\bar{p}<1$ as
\be
\mathrm{Re}\tilde{\Pi}_R(\v p,\omega) &\approx & -\frac{g}{2\pi}\frac{\bar{\omega}}{\sqrt{\bar{p}^2 - \bar{\omega}^2}} \left( 1- \frac{\bar{\omega}^2 - \bar{p}^2\sin^2\eta}{4\bar{p}^2\cos^2\eta}\right) \nn
&& \times \Theta(\bar{\omega}-\bar{p}\sin\eta) - \frac{g}{2\pi}
\ee
and
\be
\mathrm{Im}\tilde{\Pi}_R(\v p,\omega) &\approx & \frac{g}{\pi^2}\frac{\bar{\omega}}{\sqrt{\bar{p}^2 - \bar{\omega}^2}} \ln\frac{1}{16}\frac{|\bar{\omega}^2 -\bar{p}^2\sin^2\eta|}{\bp^2\cos^2\eta}. \nn
\ee
For this, we have used the identity $F(\phi,m) = K(m) - \mathrm{sn}^{-1}(z,m)$ where $K(m)$ is the complete elliptic integral of first kind and $z=\cos\phi /\sqrt{1-m\sin^2\phi}$.
The imaginary part of the inverse Jacobi elliptic function is approximated to $\mathrm{Im}[\mathrm{sn}^{-1}(x^{-1/2},1+x)] \approx -(\pi/2)(1-x/4)$ when $x \ll 1$.
At $\bar{\omega} = \bar{p}\sin\eta$, the real part of the polarizability shows a discontinuity while the imaginary part diverges logarithmically.
As $\eta$ increases to $\pi/2$, the polarizability of the nodal line semimetal becomes graphene-like, which is obvious from the definition of $\v p^\prime$ in (\ref{eq:p_prime}).

Now we discuss the low-frequency behaviors for arbitrary ${\v p}$ in the imaginary part of the polarizability.
First, when the external momentum $\v p$ is in the radial direction ($\eta = 0$), the imaginary part of the polarizability is approximated as
\be
\mathrm{Im}\tilde{\Pi}_R(\v p,\omega) &\approx & \frac{g}{2\pi^2}\left( \alpha_{\bar{p}} \bar{\omega} + \beta_{\bar{p}} \bar{\omega} \ln \bar{\omega} \right).
\ee
The coefficient $\alpha_{\bar{p}}$ for various ranges of $\bar{p}$ is given by
\be
\alpha_{\bar{p}<2} &=& \frac{4-p^2}{p(4+p^2)} \Big[c_0 -p\tan^{-1}\frac{p}{2} + \ln\frac{4-p^2}{32p^2} \Big],~~~\\
\alpha_{\bar{p}>2} &=& \frac{F\left(\phi_b,1\right) -\ln 4p}{(p/2)} -\int_{\phi_a}^{\phi_b}\frac{(4-p^2\cos^2\phi)^{\frac{1}{2}}}{p\cos\phi}d\phi, \nn
\ee
and $\alpha_{\bar{p}=2} = \ln 4$.
Here, $c_0 = \gamma_{EM} + \psi^{(0)}(1/2) = -1.38629$, $\phi_a = \cos^{-1}(2/p)$ and $\phi_b = \pi/2-\delta$ with $\delta \rightarrow 0$ where $\gamma_{EM}$ is the Euler-Mascheroni constant and $\psi^{(0)}(x)$ is the digamma function.
Also, the coefficient $\beta_{\bar{p}}$ in front of the logarithmic term is evaluated as
\be
\beta_{\bar{p}<2} &=& \frac{2(4-p^2)}{p(4+p^2)}, \\
\beta_{\bar{p}>2} &=& \frac{2}{p},
\ee
and $\beta_{\bar{p}=2} =1$.
As a result, we have logarithmic dependence ($\sim \bar{\omega} \ln \bar{\omega} $) in the imaginary part of the polarizability in the low-frequency limit.

On the other hand, if the external momentum has a finite angle ($0<\eta <\pi/2 $) with respect to the plane in which the nodal ring lies, the behaviors of the polarizability in the low-frequency limit completely change.
In this case, the imaginary part of the polarizability becomes
\be
\mathrm{Im}\tilde{\Pi}_R(\v p,\omega) &\approx & -\frac{g}{2\pi^2} \kappa_{\bar{p}}^{\eta} \bar{\omega}
\ee
where
\be
\kappa_{\bar{p}}^\eta = \int_{\phi_c}^{\frac{\pi}{2}} \left(\frac{4}{\bar{p}^2(\cos^2\eta\cos^2\phi + \sin^2\eta)} -1\right)^{\frac{1}{2}} d\phi, ~
\ee
for $\bar{p} < 2/\sin\eta$ and $\kappa_{\bar{p}}^\eta =0$ for $ \bar{p} \geq 2/\sin\eta$.
Here, $\phi_c = \cos^{-1}((4-\bar{p}^2\sin^2\eta)/\bar{p}^2\cos^2\eta )^{1/2}$.
As a result, when $0<\eta <\pi/2 $), the imaginary part of the polarizability has a simple linear dependence on the frequency if $\bar{p} < 2/\sin\eta$ and it is completely vanishing otherwise.

Finally, when $\eta = \pi/2$, the low-frequency character is exactly the same as that of graphene.
The imaginary part of the polarizability follows the power low such that it is $\sim \bar{\omega}$ for $\bar{p}<2$, $\sim \bar{\omega}^{3/2}$ for $\bar{p}=2$ and vanishes otherwise.
Those low-frequency behaviors of the imaginary part of the polarizability are summarized in the TABLE. \ref{table_1}.

\section{Plasmons}
Let us consider the RPA polarizability of the nodal line semimetal:
\be
\Pi_R^{RPA}(\v p,\omega) = \frac{\Pi_R(\v p,\omega)}{1-v_{\v p}\Pi_R(\v p,\omega)}
\ee
where $v_{\v p} = 4\pi e^2/\kappa_0 p^2$ is the Coulomb potential in the momentum space.
Both the real and imaginary parts of $\Pi_R^{RPA}(\v p,\omega)$ are drawn in Fig. \ref{fig:pol_rpa} for various external angles $\eta$.

\begin{figure}
\includegraphics[width=1\columnwidth]{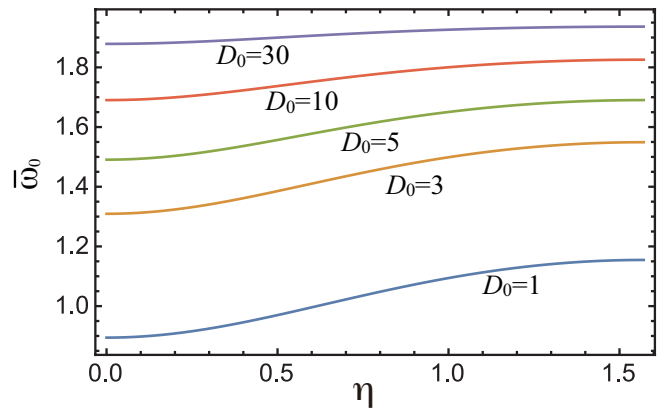}
\caption{(Color online) The plasmon frequency ($\bar{\omega}_0$) of the nodal line semimetal as a function of $\eta$ for various values of $D_0$.}
\label{fig:plasmon}
\end{figure}

The RPA polarizability $\Pi_R^{RPA}(\v p,\omega)$ shows the singularity when the dynamical dielectric function, $1-v_{\v p}\Pi_R(\v p,\omega)$ vanishes and this corresponds to the plasmon modes.
To analyze this, we obtain an approximate formula of the polarizability which is valid when $\bar{p} \rightarrow 0$ and $\bar{p} < \bar{\omega} <2$ as follows:
\be
\mathrm{Re}\tilde{\Pi}_R(\v p,\omega) &\approx & \frac{g \bar{p}^2}{8\pi} \bigg[ (1+\sin^2\eta) \frac{4-\bar{\omega}^2}{4\bar{\omega}^2} \\
&& +\frac{\bar{p}^2}{32}(3+2\sin^2\eta+3\sin^4\eta) \frac{6-\bar{\omega}^2}{\bar{\omega}^4} \bigg] \nonumber
\ee
and $\mathrm{Im}\tilde{\Pi}_R(\v p,\omega) = 0$.
Then, we obtain the plasmon mode dispersion in the long wavelength limit, described by
\be
\bar{\omega}_{pl}(p,\eta) \approx \bar{\omega}_0(\eta) + a(\eta) \bar{p}^2 \label{eq:plasmon}
\ee
where
\be
\bar{\omega}_0(\eta) &=& \left( \frac{4D_0(1+\sin^2\eta)}{4+D_0(1+\sin^2\eta)} \right)^{\frac{1}{2}}, \label{eq:omega_0}\\
a(\eta) &=& \frac{B(\eta)}{2(1+\sin^2\eta)}\frac{6-\bar{\omega}_0(\eta)^2}{\bar{\omega}_0(\eta)}
\ee
with $B(\eta) = (3+2\sin^2\eta+3\sin^4\eta)/32$ and $D_0 = g e^2 k_0/2 \kappa_0 \mu $.
The plasmon dispersion of the nodal line semimetal shares similar character with that of the conventional 3D electron gas 
in the sense that the plasma frequency is constant at $\bar{p} = 0$.
The dimensionful expression for the plasmon frequency is given by
\be
\omega_0(\eta) = \frac{\mu}{\hbar} \bar{\omega}_0(\eta) = \sqrt{\frac{4\pi}{g k_0 \hbar^2}} n^{\frac{1}{2}}_e\bar{\omega}_0(\eta).
\ee
At first glance, $\omega_0(\eta) $ appears to have the 3DEG-like $n_e^{1/2}$ dependence.  But, it has a  rather complicated dependence on the density $n_e$ and $\eta$ since $\bar{\omega}_0$ also depends on the density as shown in (\ref{eq:omega_0}).
In Fig. \ref{fig:plasmon}, we plot $\omega_0(\eta)$ as a function of $\eta$ for various values of $D_0$, which reflects the anisotropic energy spectra.
When $D_0 \gg 1$, namely when the doping is very low ($q_F/k_0 \ll e^2/\kappa_0\gamma$), the plasmon frequency becomes
\be
\omega_0(\eta) \approx 2 \frac{\mu}{\hbar} \propto n_e^{\frac{1}{2}}
\ee
which is isotropic and 3DEG-like although the quantum effect is reflected by $\hbar$.
Since the condition for the validity of the perturbative expansion, (\ref{eq:valid}), which can be rewritten roughly as $q_F/k_0 \gg (e^2/\kappa_0\gamma)^3$, one can satisfy those two conditions simultaneously when $\kappa_0 \gg 1$ which is what we usually expect from metallic systems.
On the other hand, in the opposite limit, $D_0 \ll 1$ or $q_F/k_0 \gg e^2/\kappa_0\gamma$, the plasmon frequency obeys
\be
\omega_0(\eta) \approx \frac{\mu}{\hbar}\sqrt{D_0 (1+\sin^2\eta)} \propto n^{\frac{1}{4}}_e
\ee
which is anisotropic and has the completely different power law in the electron's density.
Those are the plasmonic hallmarks of the nodal line semimetal in different regimes of the doping level.

The plasmon survives only when $\mathrm{Im}\tilde{\Pi}_R(\v p,\omega) = 0$ since the decay rate of the plasmon vanishes then.
From (\ref{eq:grp_pol_imag}), one can show that $\mathrm{Im}\tilde{\Pi}_R(\v p,\omega) = 0$ for $\bar{p}< \bar{\omega} < 2-\bar{p}$ which is the same as that of graphene.
Since $0 < \bar{\omega}_0(\eta) < 2$ for arbitrary positive value of $D_0$, the plasmon does not decay in the long wavelength limit for doped cases.

\section{Friedel Oscillations}

We start from the static polarizability, which is evaluated as
\be
\tilde{\Pi}_R(\v p,0) &=& -\frac{g}{2\pi} + \frac{g}{4\pi^2}\int_0^{\phi_s} \bar{p}^\prime G_{<}\left(2 \bar{p}^{\prime -1} \right) d\phi 
\ee
where 
\be
G_{<}(x) &=& x(1-x^2)^{1/2} - \cos^{-1}x,\\
\bar{p}^\prime &=& \bar{p}(\cos^2\eta \cos^2\phi + \sin^2\eta)^{1/2}
\ee 
and
\be
\phi_s &=& \cos^{-1}\left(\frac{\sqrt{4-\bar{p}^2\sin^2\eta}}{\bar{p}\cos\eta} \Theta\left(\frac{2}{\sin\eta} -\bar{p}\right) \right) \Theta(\bar{p}-2). \label{eq:phi_s}\nn
\ee
The static polarizability is always real valued since $\phi_s=0$ for $\bp <2$ and $2\bar{p}^{\prime -1} <1$ for $0<\phi<\phi_s$.
While $\tilde{\Pi}_R(\v p,0) = -g/2\pi$ for $\bar{p} < 2$, we only have approximate analytic formulae for $\bar{p} >2$, which are given by
\be
\tilde{\Pi}_R(\v p,0) \approx  \begin{dcases} -\frac{g}{2\pi} -\frac{g}{6\pi^2}\frac{(\bar{p}-2)^2}{\cos\eta} & (0<\eta<\frac{\pi}{2}) \\
-\frac{g}{2\pi} -\frac{g}{6\pi}(\bar{p}-2)^\frac{3}{2} & (\eta=\frac{\pi}{2}) \end{dcases} \label{eq:static_pol}
\ee
for $\bar{p} \rightarrow 2^+$ and
\be
\tilde{\Pi}_R(\v p,0) \approx -\frac{g}{8\pi}E\left( \cos^2\eta \right)\bar{p}
\ee
for $\bar{p} \gg 1$, where $E(x)$ is the complete elliptic integral.
When $\eta = \pi/2$, the polarizability is graphene-like in that its second derivative is diverging at $\bar{p} = 2$ due to the rational power law near this point.
On the other hand, when $0 <\eta < \pi/2$, this singularity is weakened so that the second derivative of the static polarizability is still singular but not diverging at $\bar{p} =2$, and just a step-like function.
The approximate formulae around $\bar{p}=2$ will be exploited in obtaining the analytic expression for the Friedel oscillations in the following.

\begin{figure}
\includegraphics[width=1\columnwidth]{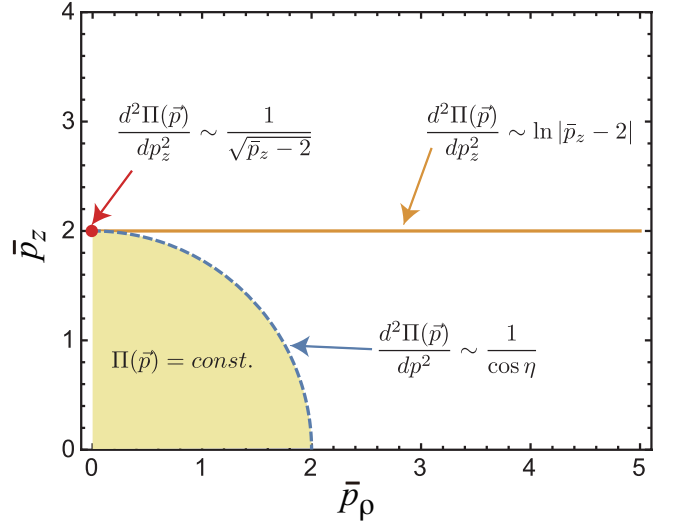}
\caption{(Color online) A schematic diagram representing the structure of singularities in the second derivative of the static polarizability. The quantity diverges along the solid line and shows a finite discontinuity (step-like) along the dashed line. In the colored region below the dashed curve, the polarizability is just a constant so that its second derivative is vanishing.}
\label{fig:static_pol}
\end{figure}

While the above-mentioned singular point ($p=2q_F$) is easily expected from usual fermionic systems, the nodal ring semimetal has another singular point at $\bar{p} = 2/\sin\eta$ ($\bar{p}_z = 2$) as indicated by the step function $\Theta (2/\sin\eta -\bar{p})$ in (\ref{eq:phi_s}).
This extra singularity can be roughly understood from the perspective of graphene since the component $\bar{p}_z = 2$ of the external momentum is the point where the 2D Dirac system at each $\phi$ in Fig. \ref{fig:coordinate} has a singular static polarizability.
However, if $\bar{p}_\rho = \bar{p}\cos\eta \neq 0$, the character of the singularity is completely different from that of graphene as shown in the following.
For this, we will use the cylindrical coordinate ($\bar{p}_\rho = \bar{p}\cos\eta$ and $\bar{p}_z = \bar{p}\sin\eta$).
We calculate the second derivative of the static polarizability because the discontinuous behavior appears from this level.
The second derivative of $\tilde{\Pi}_R(\v p,0)$ near $\bar{p}_z =2$ is given by
\be
\frac{d^2\tilde{\Pi}_R(\v p,0)}{dp_z^2} &\approx & \frac{g}{4\pi^2 p_\rho} \ln\frac{|\bar{p}_z -2|}{4} - \chi_0(\v p) \label{eq:static_pol_derivative_2}
\ee
where the continuous function $\chi_0$ is defined as
\be
\chi_0(\v p) &=& \frac{g}{4\pi^2}\bigg[ -\frac{3}{2\bar{p}_\rho} +\frac{2}{3}\frac{\bar{p}_\rho^3}{(\bar{p}_\rho^2+4)^2} \\
&& +\int_0^\frac{\pi}{2} \frac{\bar{p}_\rho^2\cos^2\phi}{(\bar{p}_\rho^2\cos^2\phi +4)^\frac{3}{2}}\cos^{-1}\frac{2}{\bar{p}_\rho^2\cos^2\phi +4} d\phi \bigg]. \nonumber
\ee
The divergence of the second derivative of $\tilde{\Pi}_R(\v p,0)$ at $\bar{p}_z =2$ is completely attributed to the first term of (\ref{eq:static_pol_derivative_2}). This logarithmic part is dominant when $\bar{p}_z \approx 2$ and it is the dominant contribution to the Friedel oscillation.
On the other hand, the divergence is too slow and $\chi_0$ is still necessary for the correct description of $d^2\tilde{\Pi}_R(\v p,0)/dp_z^2$.
This is quite similar to the usual 3D electron gas in that the polarizability begins to diverge from its second derivative.
However, the divergence is logarithmic in the nodal line semimetal while it is rational ($\sim 1/(p-2q_F)$) in the 3D electron gas.
Those results are summarized in Fig. \ref{fig:static_pol}.

\begin{figure}
\includegraphics[width=1\columnwidth]{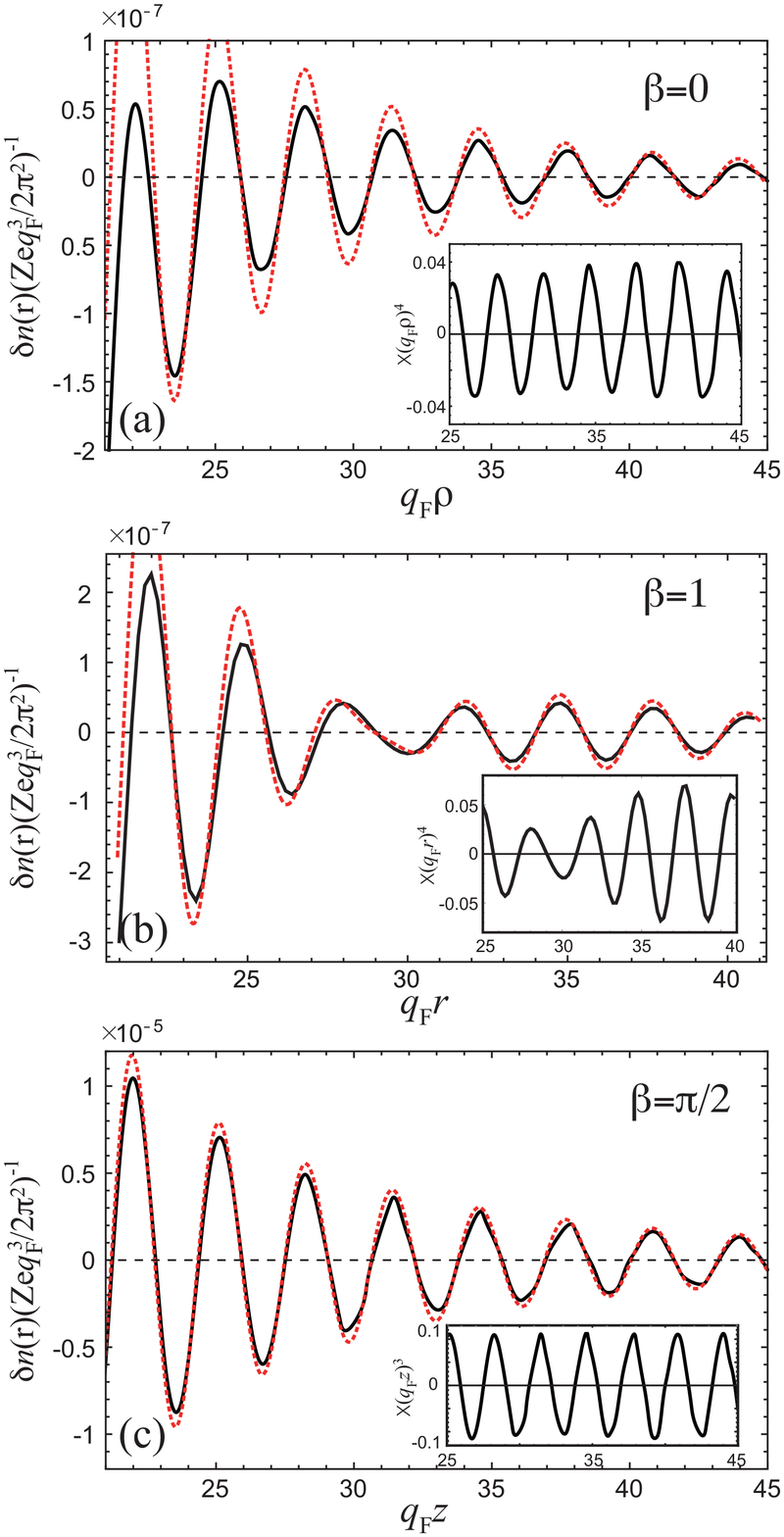}
\caption{(Color online) The Friedel oscillations of the induced charge density along (a) $\rho$, (b) $z$ and (c) $\beta=1$ directions where $\tan\beta = z/\rho$. The black solid curves are the numerical results while the red dotted ones are obtained from the approximate formulae (\ref{eq:n_I}), (\ref{eq:n_II}) and (\ref{eq:n_z}) which are valid at large distances. In each inset, the results are rescaled by $\rho^4$, $r^4$ and $z^3$. Here we assume that $\alpha =1$. }
\label{fig:friedel}
\end{figure}

When a point charge impurity $n^{ext}(\v r) = Ze\delta(\v r)$ is placed, one can evaluate the screening charge density $\delta n(\v r)$ from
\be
\delta n(\v r) = \frac{Ze}{(2\pi)^3} \int d^3p \left( \frac{1}{\epsilon (\v p)} -1\right) e^{i\v p \cdot \v r} \label{eq:induced_charge_0}
\ee
where $\v r = \rho\hat{\rho} + z\hat{z}$, and $\epsilon (\v p)$ is the static dielectric function defined by $\epsilon (\v p) = 1- v_{\v p} \tilde{\Pi}_R(\v p,0)$.
When there is a singularity at the $m$-th derivative of $\tilde{\Pi}_R(\v p,0)$, the induced charge density shows an oscillation in the real space,
which is called the Friedel oscillation.
Since we have two kinds of singularities at $\bar{p}=\sqrt{\bar{p}_\rho^2+\bar{p}_z^2} =2$ and $\bar{p}_z =2$, the charge density oscillation $\delta n(\v r)$ consists of two parts, $\delta n_I(\v r)$ and $\delta n_{II}(\v r)$ for each singularity.

First, when $q_F\rho \gg 1$, the Thomas-Fermi screening is exponentially suppressed and we obtain the approximate formulae for $\delta n_I(\v r)$ and $\delta n_{II}(\v r)$ as follows.
\be
\delta n_I(\v r) \approx \frac{Ze}{(2\pi)^3} \frac{2g\alpha \sec\beta}{3\pi\epsilon_0^2 q_F} \frac{1}{r^4}\cos 2q_F r \label{eq:n_I}
\ee
from the non-diverging singularity at $\bar{p} =2$ and
\be
\delta n_{II}(\v r) \approx \frac{Ze}{(2\pi)^3} \frac{g\alpha \sin^4\beta}{4 \epsilon_\beta^2 q_F \cos\beta} \frac{1}{r^4}\sin \frac{2q_F r}{\sin\beta} \label{eq:n_II}
\ee
from the diverging singularity at $\bar{p} = 2/\sin\eta$.
Here, $\alpha = 4\pi e^2k_0/\kappa_0\mu$,
$\tan\beta = z/\rho$, $\epsilon_0 = 1+g\alpha/8\pi$ and $\epsilon_\beta$ is the static dielectric constant at $\bar{p}=2/\sin\beta$ and $\eta=\pi/2-\beta$.
Note that $\epsilon_0 \approx \epsilon_\beta \approx 1$ when $\alpha \ll 8\pi/g$.
Both the charge density oscillations decay as $r^{-4}$.
However, they have different periodicities corresponding to the singularities at $p=2q_F$ and $p=2q_F/\sin\beta$ for a given direction from the charge impurity.
We confirm these analytic results by comparing them with the numerical results at large distances as exhibited in Fig. \ref{fig:friedel}.
Approaching the $\rho$ axis, $\delta n_I(\v r)$ dominates because the $\sin^4\beta$ factor of $\delta n_{II}(\v r) $ diminishes rapidly.
The induced charge oscillation is shown in Fig. \ref{fig:friedel}(a).
On the other hand, the Friedel oscillation along the direction with $\beta =1$ shows beat phenomenon due to the two different periodicities $\pi/q_F$ and $\pi\sin\beta/q_F$ as presented in Fig. \ref{fig:friedel}(b).
As mentioned at the beginning of this paragraph, those approximate formulae are valid for $q_F\rho \gg 1$ and not applicable to the density oscillations along $z$ axis.

Second, let us consider the Friedel oscillations along the $z$ axis ($\beta = \pi/2$).
In this case, the singularity at $\bar{p}_z = 2$ yields the charge oscillation decaying as $z^{-3}$ while the other one at $\bar{p} = 2$ gives $z^{-4}$ law.
Then, we can neglect the contribution from the singularity at $\bar{p} = 2$ far from the charge impurity and the induced charge density is given by
\be
\delta n_{II}(\v r) \approx \frac{Ze}{(2\pi)^3} \frac{\pi g\alpha }{4}\frac{\cos 2q_F z}{z^3} \label{eq:n_z}
\ee
where we assume that $\alpha \ll 8\pi/g$.
This analytic formula gives excellent agreement with the numerical calculations when $q_F z \gg 1$ as shown in Fig. \ref{fig:friedel} (c).

Consequently, the Friedel oscillation in the nodal line semimetal reflects the toroidal and anisotropic nature of its underlying band structure very well.
First, the induced charge density decays as $r^{-3}$ along the $z$ direction from the charge impurity while it obeys $r^{-4}$ law otherwise.
This reflects the difference in the strength of the singularities of the static polarizability at $\bar{p}_z =2$ and $\bar{p} =2$ as described by (\ref{eq:static_pol}) and (\ref{eq:static_pol_derivative_2}) or Fig. \ref{fig:static_pol}.
Along the $z$ direction, the behavior should be graphene-like and the $z^{-3}$ law in the charge density oscillation reflects it.
Furthermore, we have the beat phenomena in the Friedel oscillations composed of two periodicities $\pi/q_F$ and $\pi\sin\beta/q_F$ along the direction $\tan\beta = z/\rho$ from the impurity in spite of the single Fermi surface structure of the nodal line semimetal.
Since the periodicity from the singularity at $\bar{p}_z =2$ is a function of $\beta$, we have a variety of beat phenomena depending on the direction from the impurity.

\section{Conclusions}
In this study, we investigated the properties of the dynamical and static polarizability of the nodal line semimetal within the RPA.
We obtained,  in the large ring limit ($k_0 \gg q_F$), an analytic formula of the polarizability for the neutral case with the help of the results of the graphene.
For doped cases, we derived many approximate expressions in several important regions in the frequency-momentum space.
While the singular behaviors in the polarizability reflect the shape of the Fermi surface of the material, we found out that there are two singular lines of the polarizability, along $\hbar\omega = \gamma p$ and $\hbar\omega = \gamma p \sin\eta$.
This double singularity behavior is the reflection of the toroidal Fermi surface and can be understood by considering the polarizability of 2D Dirac fermions 
at each polar angle.
First, for a given momentum transfer $\v p = p_\rho \hat{\rho} + p_z\hat{z}$, the polarizability of the 2D Dirac systems around $\pm p_\rho\hat{\rho}$ contribute to the total polarizability with the same external momentum $\v p$.
On the other hand, due to the toroidal geometry, all the Dirac systems on the ring can be transferred along $z$ direction with an amount $p_z$ simultaneously.
This is why we have an extra singularity at $\hbar\omega = \gamma p \sin\eta$ in the polarizability.

We also considered the low-frequency behaviors of the imaginary part of the polarizability, which is directly related to the decay rate of the plasmon modes.
Starting from the graphene-like features at $\eta = \pi/2$, the low-frequency characteristics change drastically as $\eta$ is decreased and, finally, it shows the logarithmic dependence on the frequency at $\eta = 0$.
Using those results on the polarizability, we analyzed the unique properties of the plasmon modes and the Friedel oscillations of the doped nodal line semimetal.
The plasmon frequency is finite in the $p=0$ limit, just like the 3D electron gas.
In the low doping limit ($q_F/k_0 \ll e^2/\kappa_0\gamma$), the plasmon frequency is proportional to $n_e^{1/2}$ just like the 3DEG case.
However, as the doping level is increased, the plasmon frequency depends on the $\eta$ such that it has its minimum and maximum values at $\eta = 0$ and $\eta =\pi/2$ each due to the anisotropy of the underlying energy spectra.
In this case, the density dependence of the plasmon frequency is $\sim n_e^{1/4}$.

Finally, we studied the Friedel oscillation by evaluating the charge density induced by a point charge impurity.
As in the dynamical case, the static polarizability also has two singular paths in its second derivative due to the toroidal geometry.
They provide two completely different charge density oscillations in the real space.
The amplitude of the induced charge density oscillation ($\delta n_{I}$) from the singularity at $p= 2 q_F$ decreases as $r^{-4}$ for any direction.
On the other hand, the density oscillation ($\delta n_{II}$) from the singularity at $p_z =2 q_F$ decays as $r^{-3}$ along $z$ axis while it decays as $r^{-4}$ for $q_F \rho \gg 1$.
The difference in the algebraic power-law originates from the difference in the strength of the singularities between those two cases.
The second derivative of the static polarizability at $p_z=2q_F$ is diverging while it is just discontinuous at $p=2q_F$.
Furthermore, their periods of the oscillation are different from each other and there exist the beat phenomena when $0<\eta<\pi/2$.

\acknowledgements
We thank Y. Huh for useful discussions. This work is supported by the NSERC of Canada (YBK). YBK would like to thank the Kavli Institute for Theoretical Physics where part of this work was done. The work at KITP was supported in part by NSF Grant No. NSF PHY11- 25915.

\appendix
\section{Induced charge density}
When $\bar{\rho} = q_F \rho \gg 1$, we perform the integration (\ref{eq:induced_charge_0}) in the spherical coordinate system.
Without loss of generality, we assume that the azimuthal angle of $\bar{r}$ is zero because the nodal ring is isotropic.
Then, the induced charge density becomes
\be
\delta \tilde{n}(\v r) &=& \int d^3\bar{p} \left( \frac{1}{\epsilon (\v \bp)} -1\right) e^{i\bp\cdot \br} \\
&=& 2\pi\int d\bp d(\cos\theta) ~\bp^2 \frac{1-\epsilon(\v \bp)}{\epsilon(\v \bp)} J_0(\bp \brho \sin\theta)  \nn
&& \times \cos (\bp\bz\cos\theta) \label{eq:int_2}
\ee
where $\delta n(\v r) = (Ze q_F^3/(2\pi)^3) \delta \tilde{n}(\v r)$ and $J_0(x)$ is the Bessel function of the first kind.
$\theta$ is the polar angle of $\v \bp$ as shown in Fig. \ref{fig:coordinate}.
We use the fact that $\epsilon(\bp,\theta) = \epsilon(\bp,\pi -\theta)$ in obtaining (\ref{eq:int_2}).
If $\bar{\rho} \gg 1$, we have
\be
\delta \tilde{n}(\v r) &\approx & \sqrt{\frac{8\pi}{\brho}} \int d\bp d\theta ~\bp^\frac{3}{2}\sin^{\frac{1}{2}}\theta \frac{1-\epsilon(\v \bp)}{\epsilon(\v \bp)}  \nn
&& \times \cos\left( \bp\brho\sin\theta -\frac{\pi}{4}\right) \cos(\bp\bz\cos\theta) \\
&=& \sqrt{\frac{8\pi}{\brho}} \int d\bp d\theta ~\bp^\frac{3}{2}\sin^{\frac{1}{2}}\theta \frac{1-\epsilon(\v \bp)}{\epsilon(\v \bp)}  \nn
&& \times \cos\left( \bp \br \sin(\theta + \beta) -\frac{\pi}{4}\right)
\ee
where $\br = \sqrt{\brho^2 + \bz^2}$ and $\tan\beta = \bz / \brho$.
Here, the interval of the integration is $[0,\infty)$ for $\brho$ and $[0,\pi]$ for $\theta$.
For the integration over $\theta$, we note that only the values around $\theta = \pi/2 - \beta$ are relevant when the cosine term is highly oscillating ($\br \rightarrow \infty$) and the other terms are smooth functions.
Then, the integral is evaluated approximately as
\be
\delta \tilde{n}(\v r) &\approx & \sqrt{\frac{8\pi}{\brho}} \int d\bp~ \bp^\frac{3}{2}\sqrt{\cos\beta} \frac{1-\epsilon(\bp,\pi/2-\beta)}{\epsilon(\bp,\pi/2-\beta)} \nn
&& \times \int d\theta \cos\left( \bp \br \sin(\theta + \beta) -\frac{\pi}{4}\right) \\
&\approx & \frac{4\pi}{\br} \int d\bp~ \bp \frac{1-\epsilon(\bp,\pi/2-\beta)}{\epsilon(\bp,\pi/2-\beta)}\sin(\bp\br) \label{eq:int_6}
\ee
where we use the formula $\sin(\theta+\beta) \approx 1-(\theta+\beta-\pi/2)^2/2$ near $\theta = \pi/2-\beta$.
The approximate formula (\ref{eq:int_6}) is very accurate for $\beta \ll \pi/2$ but not bad even close to $\beta = \pi/2$.
To obtain the oscillatory part of the induced charge density, we perform the integration by part three times and only take the singular parts.
\be
\delta \tilde{n}(\v r) &\approx & \frac{4\pi}{\br^4} \int d\bp \frac{\bp \cos(\bp\br)}{\epsilon(\bp,\pi/2-\beta)} \frac{d^3 \epsilon(\bp,\pi/2-\beta)}{d\bp^3} \\
&=& \frac{4\pi}{\br^4}\Bigg[ \int d\bp \frac{g\alpha/3\pi^2}{\epsilon(\bp,\pi/2-\beta)}\frac{\delta(\bp-2)}{\bp \cos\beta}\cos(\bp\br) \nn
&& - \int d\bp \frac{g\alpha/4\pi^2}{\epsilon(\bp,\pi/2-\beta)}\frac{\sin^2\beta}{\bp^2\cos\beta}\frac{\cos(\bp\br)}{\bp-2\csc\beta} \Bigg] \nn
\ee
where we exploit two approximate formulae for the polarizability and its second derivative (\ref{eq:static_pol}) and (\ref{eq:static_pol_derivative_2}) at $\bp = 2$ and $\bp = 2\csc\beta$
for a given direction $\beta$ in the momentum space plotted in Fig. \ref{fig:static_pol}.
Finally, the integration over $\bp$ gives us
\be
\delta \tilde{n}(\v r) &\approx & \frac{2g\alpha}{3\pi\epsilon_0^2}\frac{\cos(2\br)}{\br^4\cos\beta} + \frac{g\alpha}{4\epsilon_\beta^2}\frac{\sin^4\beta}{\cos\beta}\frac{\sin(2\br\csc\beta)}{\br^4} \label{eq:int_9}
\ee
where $\epsilon_0 = 1+g\alpha/8\pi$ and $\epsilon_\beta = \epsilon(2\csc\beta,\pi/2-\beta)$.

Since (\ref{eq:int_9}) cannot be applied to the induced density oscillation along $z$ axis ($\brho = 0$), we consider this case separately.
In this case, we neglect the density oscillation from the singularity at $\bp =2$ because it decays more faster than the density oscillation from the singularity at $\bpz =2$.
Then, the induced charge density is evaluated, in the cylindrical coordinate system, as
\be
\delta \tilde{n}(\v r) &=& \int d^3\bar{p} \left( \frac{1}{\epsilon (\v \bp)} -1\right) e^{i\bp\cdot \br} \\
&=& 4\pi \int d\bpr d\bpz ~\bpr \frac{1-\epsilon(\v \bp)}{\epsilon(\v \bp)} \cos(\bpz \bz)
\ee
where $\brho =0$ is reflected and the integration range for both $\bpr$ and $\bpz$ is $[0,\infty)$.
To extract the oscillatory part, we conduct the integration by part three times which leads to
\be
\delta \tilde{n}(\v r) &\approx & -\frac{4\pi}{\bz^3} \int d\bpr d\bpz \frac{\bpr}{\epsilon(\v \bp)}\frac{d^3 \epsilon(\v \bp)}{d\bpz^3} \sin(\bpz \bz) \\
&\approx & \frac{g\alpha}{\pi\bz^3} \int d\bpr \frac{1}{\epsilon(\bpr,2)}\frac{1}{4+\bpr^2} \int d\bpz \frac{\sin(\bpz \bz)}{\bpz-2} \nn
&\approx & \frac{g\alpha}{\bz^3}\cos(2\bz)  \int d\bpr \frac{1}{\epsilon(\bpr,2)}\frac{1}{4+\bpr^2} \\
&\approx & \frac{\pi g\alpha}{4\bz^3}\cos(2\bz)
\ee
where the conditions $\bz \gg 1$ and $\alpha \ll 8\pi/g$ are applied.

\end{document}